\documentclass[fleqn,usenatbib]{mnras}

\usepackage{newtxtext,newtxmath}

\usepackage[T1]{fontenc}

\DeclareRobustCommand{\VAN}[3]{#2}
\let\VANthebibliography\thebibliography
\def\thebibliography{\DeclareRobustCommand{\VAN}[3]{##3}\VANthebibliography}

\usepackage{graphicx}	
\usepackage{amsmath}	

\title[Imprint of obliquity in pulsar spin-down]{Imprint of magnetic obliquity in apparent spin-down of radio pulsars}

\author[A. Biryukov et al.]{
Anton Biryukov$^{1,2,3}$\thanks{E-mail: ant.biryukov@gmail.com}
and Gregory Beskin$^{4,3}$
\\
$^1$Sternberg Astronomical Institute, Moscow State University, 13 Universitetsky pr., Moscow, 119234, Russia \\
$^2$Faculty of Physics, HSE University, 21/4 Staraya Basmannaya str., 
Moscow, 105066, Russia\\
$^3$Kazan Federal University, 18 Kremlyovskaya str., Kazan, 420008, Russia \\
$^4$Special Astrophysical Observatory, Nijniy Arkhyz, 
Karachai-Cherkessia, 369167, Russia
}

\date{Accepted XXX. Received YYY; in original form ZZZ}

\pubyear{2015}

\begin{document}
\label{firstpage}
\pagerange{\pageref{firstpage}--\pageref{lastpage}}
\maketitle

\begin{abstract}
Numerical simulations predict that the spin-down rate of a single rotation-powered neutron star 
depends on the angle $\alpha$ between its spin and magnetic axes as $P\dot P \propto \mu^2 (k_0 + 
k_1\sin^2\alpha)$, where $P$ is the star spin period, $\mu$ is its magnetic moment, while $k_0 
\sim k_1 \sim 1$. Here we describe a simple observational test for this prediction based on the 
comparison of spin-down rates of 50 nearly orthogonal (with $\alpha$ close to 90 deg) and
27 nearly aligned (with $\alpha$ close to 0 deg) pulsars. We found, that
the apparent pulsar spin-down is consistent with the theory if assumed, that magnetic moments of 
orthogonal rotators are systematically larger than those of aligned ones for $\sim 0.15..0.2$ 
dex.  Also, as a by-product of the analysis, we provide yet another constraint on the average 
braking index of radio pulsars as $1 \le n \le 4$ with formal significance not worse than 99\%.
\end{abstract}




\section{Introduction}
Life and observational properties of neutron stars
(NS) significantly depend on the strength and structure of their 
magnetic fields. For instance, evolution of spin period $P$ of an isolated, spherical neutron star with dipole magnetic field (a pulsar) is expected to follow the equation
\begin{equation}
    P \dfrac{dP}{dt} = \dfrac{4\pi^2 \mu^2}{Ic^3}\times (k_0 + k_1 \sin^2\alpha),
    \label{eq:spindown_law}
\end{equation}
where $\mu$ is the magnetic moment of the star and $\alpha$ is the
angle between its spin and magnetic axes -- {\it the magnetic angle} or {\it the magnetic 
obliquity} of the star. Pulsar spin-down law in this form was initially established by 
\cite{spitkovsky06} from {\it ad hoc} numerical simulations of a NS magnetosphere. The remaining 
parameters in this equation are NS moment of inertia $I$ and speed of light $c$. After, 
\cite{phil14} have estimated that dimensionless $k_0 \approx 1.0$ and $k_1 \approx 1.4$ for a pulsar with typical parameters. 

The Equation (\ref{eq:spindown_law}) is assumed to be valid for any moment within the pulsar 
lifetime even if $\mu$, $\alpha$ and $I$ are not constant. Moreover, at least two of these 
parameters are expected to be variable for real pulsars. Thus, the magnetic field should suffer 
slow decay because of the Ohmic dissipation of electric currents within the star crust 
\citep[e.g.][]{ip21}. At the same time, the same numerical calculations led to Eq. (\ref{eq:spindown_law}) have shown that $\dot\alpha < 0$ \citep{phil14}. In other words, 
the magnetic moment of an isolated pulsar must become aligned with its spin axis on evolutionary 
timescale $P/\dot P$. At the same time, the moment of inertia is unlikely to suffer any long-term 
evolution, but it changes slightly during glitches -- sudden tiny jumps in pulsar rotational
velocity \citep[e.g.][]{haskell15}.

The term in parentheses in (\ref{eq:spindown_law}) describes how
exactly the NS spins down depending on the instant orientation of its magnetosphere. 
Specifically, a nearly orthogonal pulsar (with $\alpha$ close to 90 deg) is expected to lose its 
rotational energy for $k_0 + k_1 \approx 2.4$ times faster than a nearly aligned one with the 
same $P$, $\mu$ and $I$.

In this Letter, we describe an observational test for this prediction. This test is based on 
comparing the apparent distributions of $P\dot P$  for radio pulsars with known
magnetic obliquity $\alpha$. Since the reliable estimation of $\alpha$ is 
difficult \citep[e.g][]{rankin93a, welt08, nikitina11}, we propose to analyze only pulsars with 
either nearly aligned or nearly orthogonal magnetic moments. We believe that both extreme 
states can be distinguished more confidently in observations. Indeed, light curves of orthogonal 
rotators are expected to have a narrow interpulse shifted for $\sim 180$ deg from the main one 
\citep[e.g][]{welt08}. On the other hand, $\alpha \sim 0$ deg will produce remarkably wide pulses 
in their light curves \citep[e.g][]{rankin93a, rankin93b}. Polarization properties of  
pulsars' radio emission are also meaningful for attribution them as either `aligned' or 
`orthogonal' \citep{RVM}.

To perform our analysis, we have compiled a list of 77 pulsars whose magnetic obliquity can
be found in the literature and investigated if their $P$ and $\dot P$ are agree with the spin-down law  (\ref{eq:spindown_law}). 

Below, in Sect.~\ref{sect:subset}, we describe this subset of pulsars, in
Sect.~\ref{sect:results} we provide the main results of the analysis, 
Sect.~\ref{sect:discuss} contains a short discussion about pulsar braking index
value. Conclusions are collected in Sect.~\ref{sect:conclude}.

\section{Known aligned and orthogonal pulsars}
\label{sect:subset}

\begin{table*}
 \caption{List of pulsars used in the analysis. Its first part consists of 50 nearly orthogonal pulsars with an estimated $\alpha < 16$ deg, while the second one consists of 27
 nearly aligned pulsars with $\alpha > 66$ deg (see Sect.~\ref{sect:subset} for details). Classical estimatio spin-down age $\tau_\mathrm{sd} = P/2\dot P$ are shown for each pulsar.
 The list of papers where attribution for a pulsar has been provided ($\perp$ or $\parallel$) and/or $\alpha$ has been measured is also given.}
 \label{tab:pulsars}
 \begin{tabular*}{\textwidth}{rlrl|rlrl|rlrl}
  \hline
  N & Pulsar & $\tau_\mathrm{sd}$, Myr & References & N & Pulsar & $\tau_\mathrm{sd}$, Myr & References & N & Pulsar & $\tau_\mathrm{sd}$, Myr & References\\
  \hline
1	& J0152-1637	&    10.2	&	1, 3				&   18	& J1613-5234  &    1.57	&	4				&      35	& J1843-0702 &    1.42	&	4			   \\
2	& J0406+6138	&    1.69	&	3					&   19	& J1637-4553  &    0.59	&	4, 6		    &      36	& J1849+0409 &   0.559	&	4			   \\
3	& J0525+1115	&    76.3	&	2					&   20	& J1645-0317  &    3.45	&	1, 2			&      37	& J1903+0135 &    2.87	&	1			   \\
4	& J0534+2200	& 0.00126	&	2, 4		    	&   21	& J1651-1709  &    5.08	&	1				&      38	& J1909+0007 &    2.92	&	2, 3		   \\
5	& J0627+0706	&   0.253	&	4, 5, 7		    	&   22	& J1705-1906  &    1.14	&	1, 2, 4, 6	    &      39	& J1909+1102 &     1.7	&	1			   \\
6	& J0820-1350	&    9.32	&	1					&   23	& J1713-3844  &   0.143	&	4				&      40	& J1913-0440 &    3.22	&	1			   \\
7	& J0826+2637	&    4.92	&	1, 2, 4 	    	&   24	& J1722-3207  &    11.7	&	1				&      41	& J1913+0832 &   0.466	&	4			   \\
8	& J0835-4510	&  0.0113	&	1, 2				&   25	& J1722-3712  &   0.344	&	4, 5, 6, 7	    &      42	& J1915+1410 &    96.3	&	4			   \\
9	& J0842-4851	&    1.07	&	4, 6		    	&   26	& J1731-4744 &  0.0804	&	1				&      43	& J1917+1353 &   0.428	&	2			   \\
10	& J0905-5127	&   0.221	&	4, 7				&   27	& J1739-2903 &    0.65	&	2, 4, 5, 6, 7	&      44	& J1919+0021 &    2.63	&	1, 2		   \\
11	& J0908-4913	&   0.112	&	2, 4, 5, 6, 7	    &   28	& J1748-1300 &    5.15	&	2				&      45	& J1932+2220 &  0.0398	&	2			   \\
12	& J1057-5226	&   0.535	&	1, 2, 4, 5  		&   29	& J1751-4657 &    9.06	&	1				&      46	& J1935+2025 &  0.0209	&	7			   \\
13	& J1126-6054	&    11.4	&	4, 6			    &   30	& J1813+4013 &    5.79	&	1				&      47	& J2022+2854 &    2.87	&	1, 2		   \\
14	& J1413-6307	&   0.842	&	4, 6		    	&   31	& J1820-0427 &     1.5	&	2				&      48	& J2047+5029 &    1.69	&	4			   \\
15	& J1509+5531	&    2.34	&	1				   	&   32	& J1828-1101 &  0.0772	&	4, 5, 7	    	&      49	& J2155-3118 &    13.2	&	1			 	\\
16	& J1549-4848	&   0.324	&	4, 6, 5, 7	    	&   33	& J1841+0912 &    5.54	&	1, 3			&      50	& J2330-2005 &    5.62	&	2			   \\
17	& J1611-5209	&   0.559	&	4					&	34	& J1842+0358 &    4.56	&	4	 			&			&			 &			& 					\\

  \hline
   \\
  \hline
1	& J0157+6212 &	 0.197 &  2		    &  10	& J1424-6438 &	  67.6	&  4, 6     &    19	& J1903+0925 &	 0.153	&  4, 6	 \\
2	& J0502+4654 &	  1.81&  1, 3		&  11	& J1543+0929 &	  27.4	&  1, 2     &    20	& J1910+0358 &	  8.26	&  1, 2	 \\
3	& J0659+1414 &	 0.111&  1			&  12	& J1637-4450 &	  6.96	&  4, 6     &    21	& J1921+1948 &	  14.5	&  1		 \\
4	& J0828-3417 &	  29.4&  1, 2, 4, 6	&  13	& J1720-0212 &	  91.4	&  1	    &    22	& J1926+1434 &	  95.6	&  1		 \\
5	& J0831-4406 &	  3.86&  4, 6		&  14	& J1806-1920 &	   819	&  4, 6     &    23	& J1946+1805 &	   290	&  4, 6	 \\
6	& J0946+0951 &	  4.98&  1			&  15	& J1808-1726 &	   329	&  4, 6	 	&    24	& J2006-0807 &	   200	&  1, 3	 \\
7	& J0953+0755 &	  17.5&  1, 4, 6	&  16	& J1834-0426 &	  63.9	&  1, 2, 3	&    25	& J2113+4644 &	  22.5	&  1, 2	 	\\
8	& J1057-5226 &	 0.535&  6			&  17	& J1851+0418 &	  4.14	&  4, 6	 	&    26	& J2149+6329 &	  35.8	&  1		 	\\
9	& J1302-6350 &	 0.332&  4			&  18	& J1852-0118 &	  4.07	&  4, 6	 	&    27	& J2325+6316 &	  8.05	&  1		 \\
\hline
 \end{tabular*}
 References legend: [1] \cite{lyne88}; [2] \cite{rankin90}; [3] \cite{rankin93a}; 
 [4] \cite{maciesiak11a}; [5] \cite{keith10}; [6] \cite{malov13}; [7] \cite{jk19}
\end{table*}

The subset of pulsars we analyze in our work consists of 50 nearly orthogonal and 27 nearly 
aligned rotators (see Table~\ref{tab:pulsars}). Hereafter we will mark all the quantities related 
to them by subscripts ``$\perp$'' and ``$\parallel$'' respectively. The pulsars were attributed 
to one of two types mainly accordingly to the values of their magnetic angles obtained from the 
literature. Specifically, we adopted estimations of $\alpha$ from \cite{lyne88}, 
\cite{keith10} and \cite{jk19}, where polarization properties of pulsars have 
been studied in the framework of the Rotation Vector Model \citep{RVM}. We also used the 
results published by \cite{rankin90, rankin93a} after the analysis of pulsars' pulse profiles' 
shapes and widths. Finally, qualitative attribution made by \cite{maciesiak11a} and 
\cite{malov13}, who focused on the interpulses, have been also taken into account. 

In a sense of spin-down rate, pulsars can be technically sorted between two types as 
follows. Let the apparent $|\dot P|$ of a pulsar be less than 10\% faster than that for a 
completely aligned rotator ($\alpha = 0$) with the same parameters. Such a pulsar can be 
considered a ``nearly aligned'' one. Numerically it means, that
\begin{equation}
    k_0 + k_1\sin^2\alpha_{\parallel} < 1.1k_0,
\end{equation}
or $\alpha_\parallel \lesssim 16^\circ$ assuming $k_0 = 1$ and $k_1 = 1.4$. The same criterion can be also applied for ``nearly orthogonal'' objects so that
\begin{equation}
    k_0 + k_1\sin^2\alpha_{\perp} > 0.9(k_0 + k_1)
\end{equation}
and hence $\alpha_\perp \gtrsim 66^\circ$. Thus, $\alpha_{\parallel,\mathrm{max}} = 
16^\circ$ and $\alpha_{\perp,\mathrm{min}} = 66^\circ$ could be, in principle, used as thresholds 
in the classification procedure. A fine point here is that actual magnetic obliquities are known 
with precision worse than just a few degrees \cite[e.g.][]{manch10} So, thresholds that are given 
with 1-degree precision are obviously over-accurate. Therefore, conclusions based on these values 
are not irreversible. And, when possible, have to be supported by additional estimations of 
$\alpha$ or other complementary arguments. In our subset, 39 of 77 pulsars have a consistent 
classification made accordingly to at least two sources.\footnote{Notice, also, that nearly a 
dozen pulsars have been excluded from further analysis since their classification was 
contradictory when based on different papers.} 

\begin{figure}
 \includegraphics[width=\columnwidth]{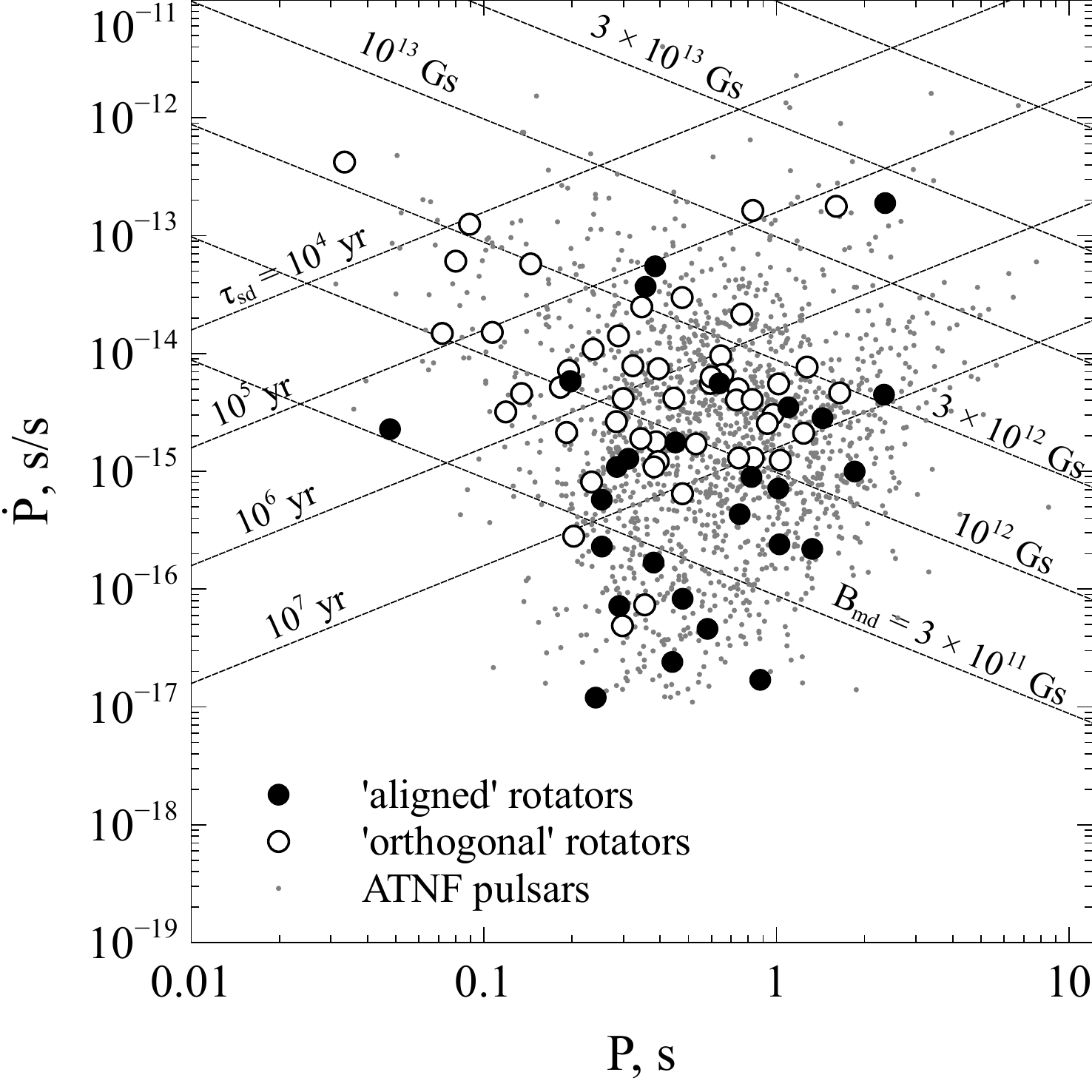}
 \caption{Period - period derivative diagram for pulsars under analysis. Nearly orthogonal 
 pulsars are shown by black  circles, while nearly aligned by white ones. Dots represent single classical pulsars from the 
 ANTF catalogue. Constant levels of conventional estimators are also shown: spin-down age $\tau_\mathrm{sd} = 
 P/2{\dot P}$ and surface magnetic field $B_\mathrm{md} \propto \sqrt{P {\dot P}}$ respectively.}
 \label{fig:ppdot}
\end{figure}

Table~\ref{tab:pulsars} contains the details of the final pulsars' subset, while the $P-\dot P$ 
diagram for it is shown in  Figure~\ref{fig:ppdot}. Other single classical (non-recycled) pulsars 
taken from the ANTF catalogue 
\citep{atnf}\footnote{\texttt{https://www.atnf.csiro.au/research/pulsar/psrcat/}, ver. 1.68
} are also represented in this plot. It appears that orthogonal pulsars have 
systematically lower spin-down ages $\tau_\mathrm{sd} = P/2\dot P$: $\langle 
\tau_\mathrm{sd,\perp} \rangle \approx 2$ Myr versus $\langle 
\tau_\mathrm{sd,\parallel} 
\rangle \approx 79$ Myr. In this sense, they are younger than the aligned ones. 
For instance, there are 19 orthogonal pulsars with  $\tau_\mathrm{sd} < 1$ Myr, but only 5 
aligned pulsars with the same age boundary. Moreover, there are no nearly aligned pulsars younger 
than  $\tau_\mathrm{sd} = 100$ kyr at all. This is consistent with the idea of evolutionary 
decreasing of magnetic angle $\alpha$, derived theoretically \citep{phil14} and after 
statistical analysis of magnetic angles distribution \citep{tm98}. Although, it can not be a 
crucial argument in favour of it \citep[e.g][]{Novoselov2020}.

\begin{figure}
 \includegraphics[width=\columnwidth]{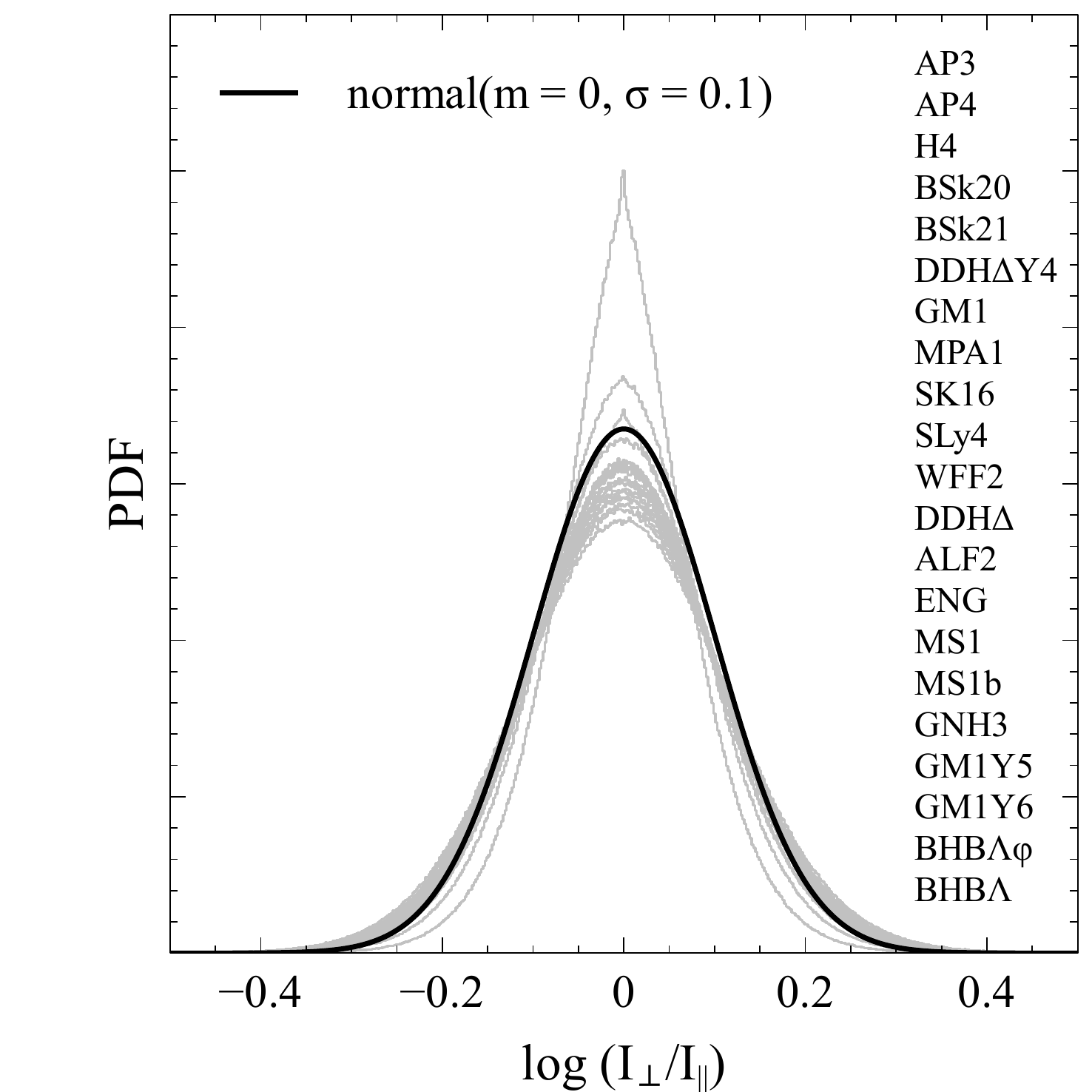}
 \caption{Distribution of the ratio $\log I_\perp/I_\parallel$ for NSs moments of inertia. Values 
 of $I_\perp(M)$ and $I_\parallel(M)$ were calculated as 
 moments of inertia of two independent pulsars with randomly distributed masses adopting one of 
 the equations of state (see text for details).
 Generally, $\log I_\perp/I_\parallel$ can be approximated by a Gaussian law with zero average
 and standard deviation of 0.1.}
 \label{fig:kcoeff}
\end{figure}

\section{Results}
\label{sect:results}

\begin{figure}
 \includegraphics[width=\columnwidth]{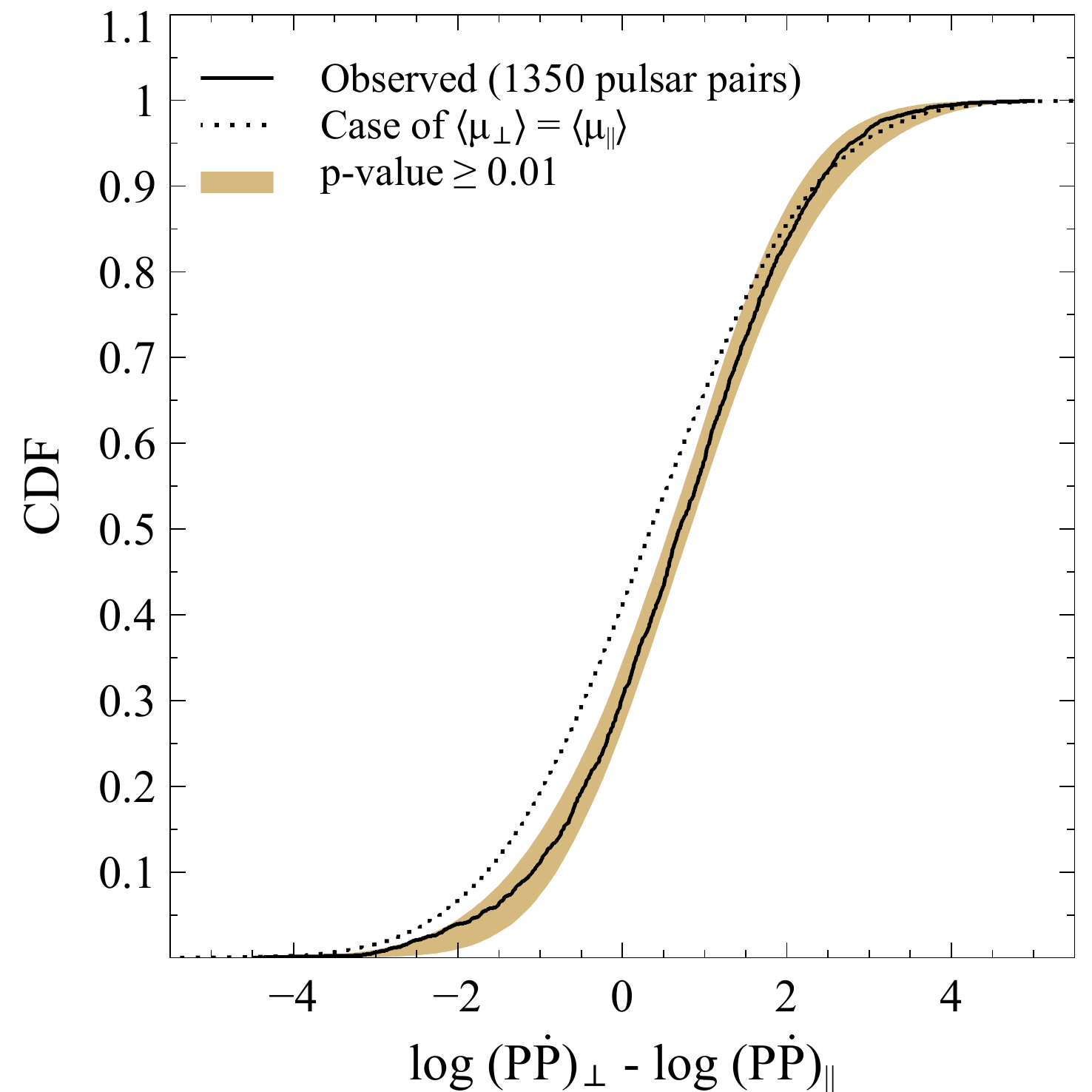}
 \caption{Black line: observed cumulative distribution of logarithmic difference $x_\mathrm{sd}$} for 1350 pairs of orthogonal and aligned pulsars. Black dotted line: theoretical distribution of $x_\mathrm{sd}$ obtained under the assumption of non-evolving magnetic fields (specifically, $\langle\mu_\perp\rangle = \langle\mu_\parallel\rangle$, $\sigma_\mu = 0.55$).
 This distribution is inconsistent with the observed one at $3\times 10^{-14}$ significance level according to the two-sample
 Kolmogorov-Smirnov test. On the other hand, the orange area around the observed curve is covering all theoretical distributions 
 that assume various $\langle\log(\mu_\perp/\mu_\parallel)\rangle \neq 0$ and which are statistically consistent with the observed one at better than 99\% significance (see Figure \ref{fig:mu_evol}).
 \label{fig:result}
\end{figure}

We compare the spin-down of orthogonal and aligned pulsars by calculating the
the cumulative distribution function for the logarithmic difference
\begin{equation}
    x_\mathrm{sd} = \log (P\dot P)_\perp - \log (P\dot P)_\parallel,
    \label{eq:xsd}
\end{equation}
where $(P\dot P)_\perp$ is taken for a nearly orthogonal pulsar and
$(P\dot P)_\parallel$ for a nearly aligned one. There are 27$\times$50 = 1350 
of such independent pairs based on our data set. 
Apparent distribution of $x_\mathrm{sd}$ is shown in 
Figure~\ref{fig:result} by a black line. It average
$\langle x_{\rm sd,obs} \rangle \approx 0.655 \pm 0.037$. In other words
orthogonal pulsars spin down $\sim 4.5$ times faster than aligned ones.

On the other hand, theoretical distributions for $x_\mathrm{sd}$
were also calculated assuming various model parameters.
From (\ref{eq:spindown_law}) one can rewrite $x_\mathrm{sd}$ as
\begin{equation}
    x_{\rm sd} = -\log \left( \dfrac{I_\perp}{I_\parallel} \right) + 2\log\left( \dfrac{\mu_\perp}{\mu_\parallel} \right) + \log \left ( \dfrac{k_0 + k_1\sin^2\alpha_\perp}{k_0 + k_1\sin^2\alpha_\parallel} \right).
    \label{eq:xsd_theoretical}
\end{equation}
Thus, this quantity is a sum of three terms, each of which we assumed to be
statistically independent on the other two\footnote{There is no fundamental reason, however, why the moment of inertia, magnetic field, its orientation and spin-down parameters ($k_0$, $k_1$) could not depend on each other.}. And each of them was calculated as follows. First, uniformly distributed angles 
$\alpha_\perp$ and $\alpha_\parallel$ were generated so that 
$\alpha_\perp \sim {\rm uniform}(0, 16^\circ)$ and $\alpha_\perp \sim {\rm uniform}(66^\circ, 
90^\circ)$ respectively. These values are then used to calculate the last term of 
(\ref{eq:xsd_theoretical}). The second term of (\ref{eq:xsd_theoretical}) depends on the assumed 
pulsar magnetic field distribution. Following the \cite{fgk06} we adopted normally distributed
magnetic fields:
\begin{equation}
    \log \left(\dfrac{\mu_\perp}{\mu_\parallel}\right) \sim \mathrm{normal}(\langle \log(\mu_\perp/\mu_\parallel) \rangle, \sigma_\mu\sqrt{2}).
    \label{eq:xmu}
\end{equation}
If $\langle \log(\mu_\perp/\mu_\parallel) \rangle = 0$, then no
difference between magnetic field distributions of two types
of pulsars is assumed.

The computation of $\log(I_\perp/I_\parallel)$ is more sophisticated. Thus, firstly we assumed 
that moment of inertia of a neutron star is independent of both pulsar obliquity and 
magnetic field. Therefore $I_\perp/I_\parallel$ follows the same distribution as the ratio of 
moments of inertia of just two randomly taken pulsars.

Assuming that masses of neutron stars distributed normally with average $1.49M_\odot$ and 
standard deviation $0.19M_\odot$ \citep{ozel16}, and adopting various equation of 
state \cite[see][for details]{bab17} we have calculated distribution of 
$\log(I_\parallel/I_\perp)$ explicitly for each EoS. The result is shown in 
Figure~\ref{fig:kcoeff}  We conclude, that this distribution can be, in general, described 
as a normal one with zero average and a standard deviation of 0.1. 

\begin{figure}
 \includegraphics[width=\columnwidth]{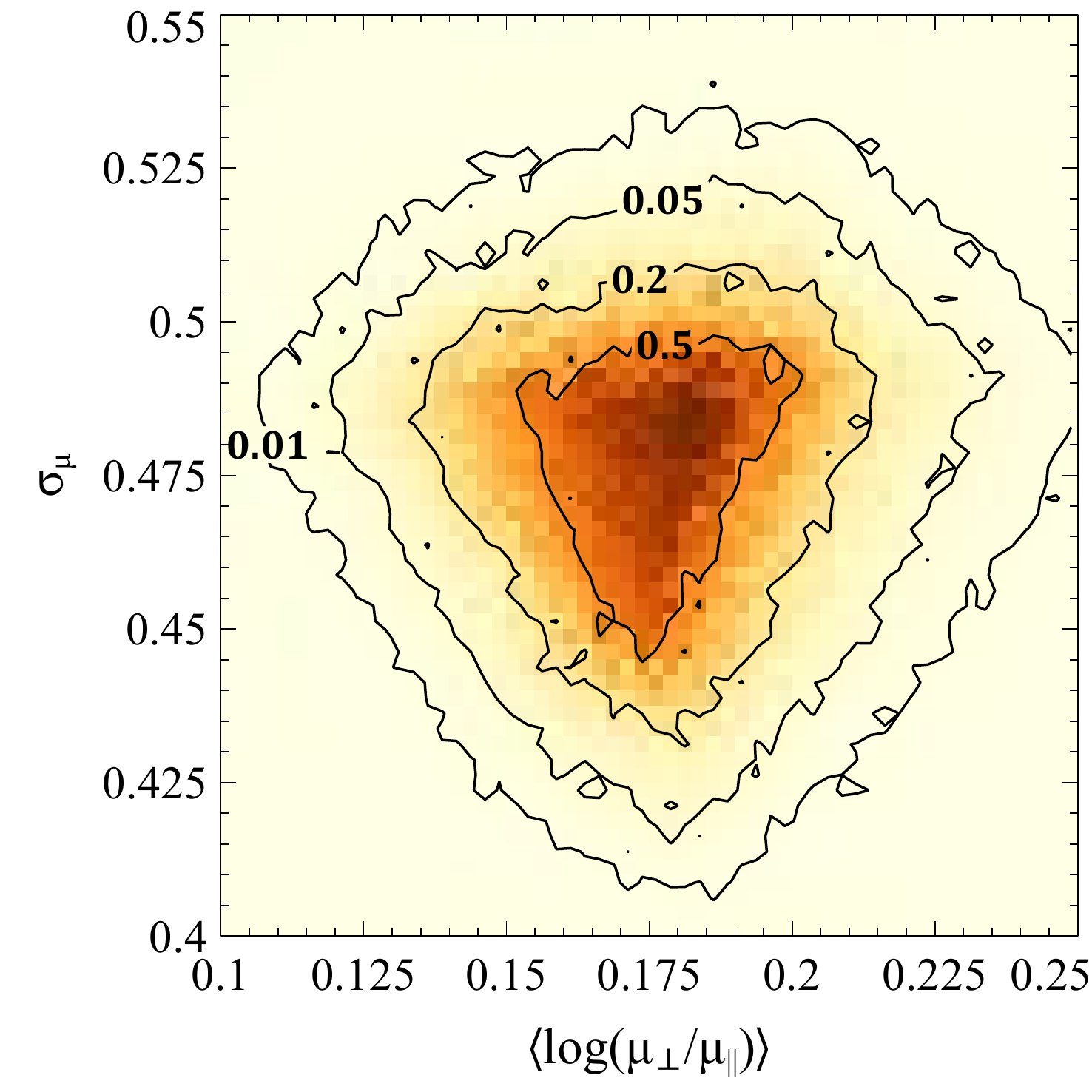}
 \caption{Distribution of KS-test p-values calculated for pairs of 
 observed and theoretical distributions of $x_\mathrm{sd}$, assuming
 that magnetic field distributions of orthogonal and aligned pulsars are not the same but $\log(\mu_\perp/\mu_\parallel) \sim \mathrm{normal}(\langle \log(\mu_\perp/\mu_\parallel) \rangle,\sigma_\mu\sqrt{2})$. Black contours bound the areas where KS p-value is larger than 0.01, 0.05, 0.2 and 0.5 respectively. {It is clear, that orthogonal pulsars must have $\approx 0.12..0.22$ dex larger magnetic moments to be consistent with the observations.}}
 \label{fig:mu_evol}
\end{figure}

At last, the theoretical value of $x_{\rm sd}$ was generated as a 
sum of three aforementioned terms. Its cumulative distribution is 
shown in Figure~\ref{fig:result} by a dotted line (for case
of zero $\langle \log(\mu_\perp/\mu_\parallel) \rangle$ and $\sigma_\mu 
= 0.55$) of by orange area around the observed CDF. This area covers
all theoretical CDFs which are consistent statistically with the 
observed one. For them we adopted various $\langle 
\log(\mu_\perp/\mu_\parallel) \rangle$ and $\sigma_\mu$ and calculated
Kolmogorov-Smirnov two-sample p-value at each step. The distribution of 
all p-values is shown in Figure \ref{fig:mu_evol}. The 
contours  that bound areas
with p-value larger than 0.01, 0.05, 0.2 and 0.5 correspondingly
are also shown. If the p-value is larger than 0.01, then the model can be
considered statistically consistent with the observations at better
than 99\% significance.

One can see, that case of non-evolving magnetic fields is far beyond
the obtained credible area of $\log(\mu_\perp/\mu_\parallel)$ distribution parameters. Indeed, KS-
test gives the probability $p \approx 3\times 10^{-14}$ that observed and synthetic subsets are 
statistically consistent.

Instead, pulsar data seem to be consistent with the assumption that 
orthogonal pulsars have systematically larger magnetic fields (for $\approx 0.17$ dex), 
while the standard  deviation for $\sigma_\mu$ over pulsar population is $\approx 0.47$ dex. The 
latter value is consistent with $\sigma_\mu \approx 0.55$ got from population synthesis 
\citep{fgk06, Igoshev22}. And technically consistent with the idea of decaying magnetic fields of 
isolated neutron stars \citep[e.g.][]{vigano13, ip15}, if one assumes that aligned rotators are 
systematically older than the orthogonal ones.

Therefore, the general spin-down law (\ref{eq:spindown_law}) seems 
consistent with the statistics of aligned and orthogonal pulsars' spin-down, if 
the appropriate difference in magnetic fields of two types of pulsars is assumed. 
This is the main result of our work.

\section{Discussion: pulsar braking index}
\label{sect:discuss}

The results described above were obtained assuming that Equation (\ref{eq:spindown_law}) is valid 
for any pulsar at every moment. This law has a purely theoretical origin and is supported by 
detailed numerical simulations. It is widely accepted as long as it is highly successful in 
reproducing statistics of pulsar timing parameters in the framework of population synthesis 
\citep{fgk06, ridley10, gullon14}. However, no direct observational evidence for 
(\ref{eq:spindown_law}) have been obtained so far. 

One of the long-standing, key problems in pulsar spin-down physics is the value of the so-called 
{\it braking index}. Let's assume, that 
pulsar spin-down can be described in general by the equation
\begin{equation}
    {\dot P}P^{n-2} = f(\mu, I, \alpha),
    \label{eq:spindown_general}
\end{equation}
where $f(\cdot)$ does not depend on spin period $P$ explicitly. The parameter $n$ here is the 
braking index -- a dimensionless quantity, which value reflects spin-down physics. The physical 
sense of $n$ became more clear when Equation (\ref{eq:spindown_general}) is 
rewritten in terms of the spin frequency $\Omega = 2\pi/P$ as $\dot\Omega \propto \Omega^n$.

The standard model (\ref{eq:spindown_law}) corresponds to $n = 3$ -- the canonical value. However,
the actual value of $n$ is still poorly constrained by the observations. In principle, it can be 
estimated as $n_\mathrm{obs} = \Omega 
\ddot\Omega/{\dot \Omega}^2$ assuming that $f(\cdot)$ is constant 
for a given pulsar. But, as long as $\ddot \Omega$ for real objects is strongly affected by 
neutron star rotational irregularities, 
$n_\mathrm{obs}$ is confidently known only for a handful of objects \citep[see e.g.][and 
references therein]{hobbs10, bbk12}. However,
even in these cases, $n$ shows a pretty wide range of values from $\sim 0.9$ to $3.15$ 
\citep{archi16}. Therefore, any observational test of the true value of $n$ still makes sense.

The analysis of spin periods and their derivatives of orthogonal and aligned pulsars avoid using 
$\ddot\Omega$ (to $\ddot P$). This allowed us to suggest a method for the rough constraint of $n$ 
using the data set discussed in this work. The idea of the method is as follows. Let's introduce 
a more general spin-down ratio based on (\ref{eq:spindown_general}):
\begin{equation}
	x_\mathrm{sd}(n) = \log (P^{n-2}\dot P)_\perp - \log (P^{n-2}\dot P)_\parallel.
    \label{eq:xsd_n}
\end{equation}
Theoretically, the distribution of this quantity should be the same as that of the 
logarithmic ratio
\begin{equation}
	y = \log \left(\dfrac{f(\mu_\perp, I, \alpha_\perp)} {f(\mu_\parallel, I, \alpha_\parallel)} \right ) = \log f_\perp - \log f_\parallel
\end{equation}
if $n$ is equal to the true value. On the other hand, the distribution of $\log(f_\perp/f_\parallel)$ 
can be independently estimated from the distribution of $x_\mathrm{sd} = \log({\dot P 
P}_\perp/{\dot P P}_\parallel)$ taken only for pairs of pulsars which 
periods are either equal or very close to each other. A small difference in $P_\perp$ and 
$P_\parallel$ in this case making $x_\mathrm{sd}$ technically close to $\log({\dot P}_\perp/{\dot 
P}_\parallel)$ which does not depend on $n$ explicitly.

\begin{figure}
 \includegraphics[width=\columnwidth]{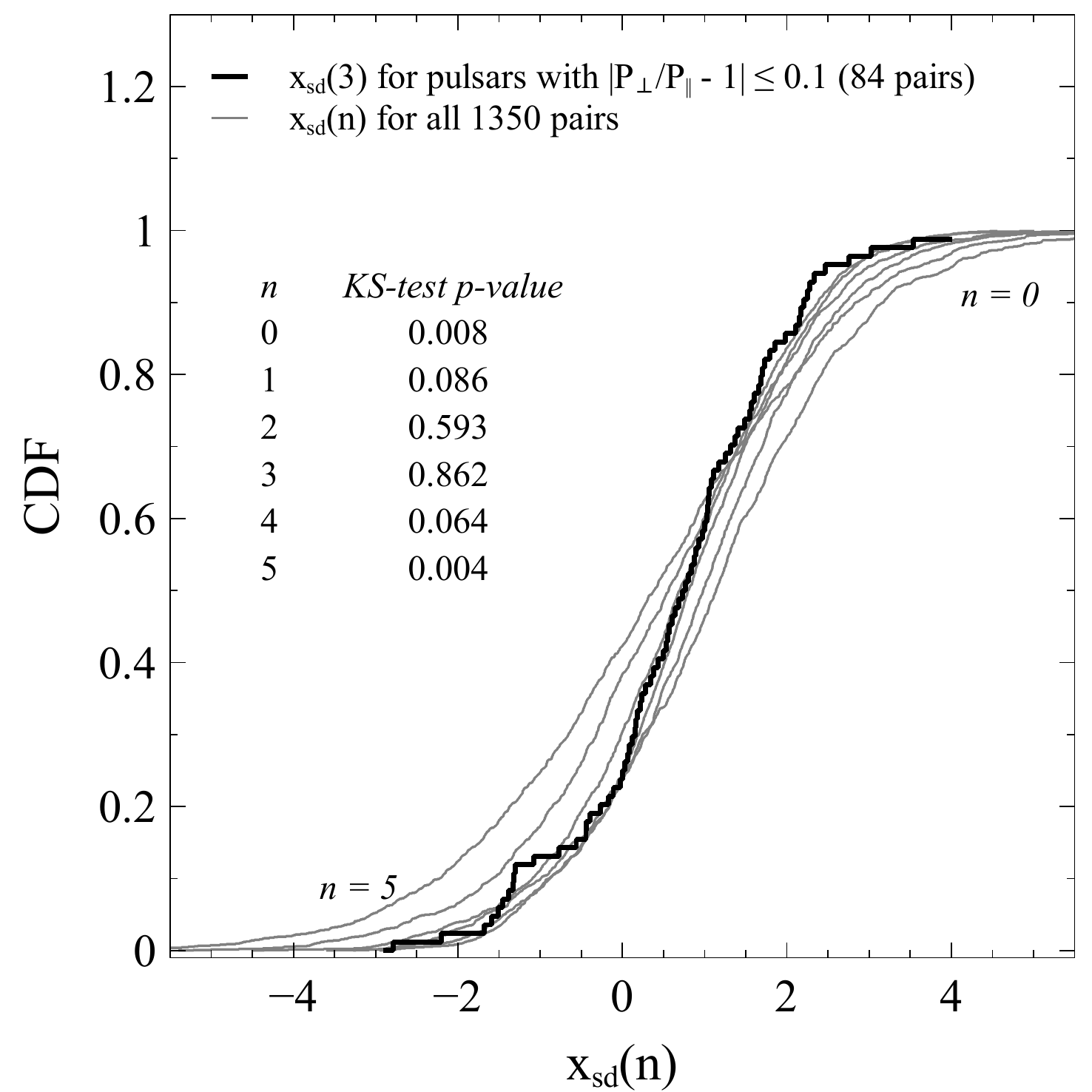}
 \caption{Grey lines: CDF of $x_\mathrm{sd}(n)$ (Eq.~\ref{eq:xsd_n}) calculated over 
 1350 pairs of pulsars for $n = 0..5$. Black line: CDF of $x_\mathrm{sd}$ for 84 pairs of 
 pulsars with close periods. These distributions are statistically equal while
 $n = 1..4$ which gives simple constraint on value of braking index
 in the pulsar spin-down law.}
 \label{fig:bi}
\end{figure}

Thus, from the subset of pulsars under investigation, we have formed 84 pairs of objects such 
that $|P_\perp/P_\parallel - 1| < 0.1$. Distribution of $x_\mathrm{sd}$ for them is shown in 
Figure~\ref{fig:bi} by the thick black line. 
We compare it with the distribution of $x_\mathrm{sd}(n)$ calculated for all 1350 pulsar pairs 
assuming $n = 0..5$ with step $1$.
It has been found, that corresponding KS-test p-values remain larger 
than $0.01$ for $n$ in 1...4. This interval can be considered as a simple constraint on the value of braking index $n$ in (\ref{eq:spindown_general}) at significance not worse than $99\%$.

\section{Conclusions}
\label{sect:conclude}

The undertaken analysis has shown that:
\begin{itemize}
    \item observed statistics of $P\dot P$ of nearly aligned and nearly
    orthogonal pulsars is consistent with theoretical spin down law (\ref{eq:spindown_law}) obtained from 3D simulations;
    \item magnetic fields of orthogonal pulsars are $\sim 0.2$ dex stronger than that of aligned ones;
    \item if one assumes pulsar spin-down law in the more general form $\dot P P^{2-n} = f(\mu, I, \alpha)$, then $n$ lies within 1..4 with confidence not worse 99\%;
\end{itemize}

The second conclusion is qualitatively consistent with the idea of magnetic field decay. Indeed, 
if orthogonal pulsars in our subset 
are younger than the aligned ones, then pulsar magnetic fields decay for $\sim 0.2$
dex during tens of Myr of pulsar evolution. This is quite a slow field evolution
rate, but this is consistent with that obtained for more sophisticated 
analysis in other works \citep[e.g. see][and references therein]{ip14, ip15, ip21}.

On the other hand, as it has been argued in \cite{Novoselov2020}, 
observed orthogonal pulsars are expected to have larger magnetic fields
because the pulsar ``death line'' condition depends on the obliquity $\alpha$.
Specifically, an orthogonal pulsar with a relatively weak magnetic field
crosses the death line earlier than the aligned one with the same parameters.
Therefore, the magnetic field decay effect seen in our analysis could be partially due to 
observational selection effects. It can be accurately taken into account in the framework of 
single pulsar population synthesis, which is beyond the scope of our work.

It could be only noticed, that there have been many attempts to reproduce actual statistics of 
pulsar parameters with synthetic distributions. However, no solid conclusion about  the empirical 
magnetic evolution model has been made. Moreover, it remains unclear whether magnetic 
evolution is really necessary for reproducing properties of the pulsar population in hand. 

The problem there are conflicting results obtained within various population 
synthesis runs. Namely, the properties of isolated  rotation-powered pulsars 
were satisfactorily reproduced within  different models of pulsar 
evolution: by assuming either constant magnetic fields \citep{fgk06, ridley10} or slowly decaying fields \citep{bha92, popov10} or, vice versa, rapidly decaying fields \citep{lmt85, gullon14}. These paradoxical results probably indicate that population synthesis of normal 
pulsars is, in fact, weakly sensitive to the assumption about NS magnetic history. 

Thus, we have to admit, that magnetic field decay could be the reliable, but not the ultimate interpretation of the discrepancy between observed and modelled distributions of pulsars $\log(P\dot P)_\perp - \log(P \dot P)_\parallel$. Further investigation of this effect is still needed.

\section*{Acknowledgment}
This work was supported under the Ministry of Science and
Higher Education of the Russian Federation grant 075-15-2022-262
(13.MNPMU.21.0003).

\section*{Data Availability}
The data underlying this article will be shared on reasonable request to the corresponding author.


\bibliographystyle{mnras}
\bibliography{eman} 

\begin{thebibliography}{}
\makeatletter
\relax
\def\mn@urlcharsother{\let\do\@makeother \do\$\do\&\do\#\do\^\do\_\do\%\do\~}
\def\mn@doi{\begingroup\mn@urlcharsother \@ifnextchar [ {\mn@doi@}
  {\mn@doi@[]}}
\def\mn@doi@[#1]#2{\def\@tempa{#1}\ifx\@tempa\@empty \href
  {http://dx.doi.org/#2} {doi:#2}\else \href {http://dx.doi.org/#2} {#1}\fi
  \endgroup}
\def\mn@eprint#1#2{\mn@eprint@#1:#2::\@nil}
\def\mn@eprint@arXiv#1{\href {http://arxiv.org/abs/#1} {{\tt arXiv:#1}}}
\def\mn@eprint@dblp#1{\href {http://dblp.uni-trier.de/rec/bibtex/#1.xml}
  {dblp:#1}}
\def\mn@eprint@#1:#2:#3:#4\@nil{\def\@tempa {#1}\def\@tempb {#2}\def\@tempc
  {#3}\ifx \@tempc \@empty \let \@tempc \@tempb \let \@tempb \@tempa \fi \ifx
  \@tempb \@empty \def\@tempb {arXiv}\fi \@ifundefined
  {mn@eprint@\@tempb}{\@tempb:\@tempc}{\expandafter \expandafter \csname
  mn@eprint@\@tempb\endcsname \expandafter{\@tempc}}}

\bibitem[\protect\citeauthoryear{{Archibald} et~al.,}{{Archibald}
  et~al.}{2016}]{archi16}
{Archibald} R.~F.,  et~al., 2016, \mn@doi [Astrophys. J. Lett.]
  {10.3847/2041-8205/819/1/L16}, \href
  {http://adsabs.harvard.edu/abs/2016ApJ...819L..16A} {819, L16}

\bibitem[\protect\citeauthoryear{{Bhattacharya}, {Wijers}, {Hartman}  \&
  {Verbunt}}{{Bhattacharya} et~al.}{1992}]{bha92}
{Bhattacharya} D.,  {Wijers} R.~A.~M.~J.,  {Hartman} J.~W.,   {Verbunt} F.,
  1992, Astron. and Astrophys., \href
  {http://adsabs.harvard.edu/abs/1992A%26A...254..198B} {254, 198}

\bibitem[\protect\citeauthoryear{{Biryukov}, {Beskin}  \& {Karpov}}{{Biryukov}
  et~al.}{2012}]{bbk12}
{Biryukov} A.,  {Beskin} G.,   {Karpov} S.,  2012, \mn@doi [MNRAS]
  {10.1111/j.1365-2966.2011.20005.x}, \href
  {http://adsabs.harvard.edu/abs/2012MNRAS.420..103B} {420, 103}

\bibitem[\protect\citeauthoryear{{Biryukov}, {Astashenok}  \&
  {Beskin}}{{Biryukov} et~al.}{2017}]{bab17}
{Biryukov} A.,  {Astashenok} A.,   {Beskin} G.,  2017, \mn@doi [MNRAS]
  {10.1093/mnras/stw3341}, 466, 4320

\bibitem[\protect\citeauthoryear{{Faucher-Gigu{\`e}re} \&
  {Kaspi}}{{Faucher-Gigu{\`e}re} \& {Kaspi}}{2006}]{fgk06}
{Faucher-Gigu{\`e}re} C.-A.,  {Kaspi} V.~M.,  2006, \mn@doi [Astrophys. J.]
  {10.1086/501516}, \href {http://adsabs.harvard.edu/abs/2006ApJ...643..332F}
  {643, 332}

\bibitem[\protect\citeauthoryear{{Gull{\'o}n}, {Miralles}, {Vigan{\`o}}  \&
  {Pons}}{{Gull{\'o}n} et~al.}{2014}]{gullon14}
{Gull{\'o}n} M.,  {Miralles} J.~A.,  {Vigan{\`o}} D.,   {Pons} J.~A.,  2014,
  \mn@doi [MNRAS] {10.1093/mnras/stu1253}, \href
  {http://adsabs.harvard.edu/abs/2014MNRAS.443.1891G} {443, 1891}

\bibitem[\protect\citeauthoryear{{Haskell} \& {Melatos}}{{Haskell} \&
  {Melatos}}{2015}]{haskell15}
{Haskell} B.,  {Melatos} A.,  2015, \mn@doi [International Journal of Modern
  Physics D] {10.1142/S0218271815300086}, \href
  {https://ui.adsabs.harvard.edu/abs/2015IJMPD..2430008H} {24, 1530008}

\bibitem[\protect\citeauthoryear{{Hobbs}, {Lyne}  \& {Kramer}}{{Hobbs}
  et~al.}{2010}]{hobbs10}
{Hobbs} G.,  {Lyne} A.~G.,   {Kramer} M.,  2010, \mn@doi [MNRAS]
  {10.1111/j.1365-2966.2009.15938.x}, \href
  {http://adsabs.harvard.edu/abs/2010MNRAS.402.1027H} {402, 1027}

\bibitem[\protect\citeauthoryear{{Igoshev} \& {Popov}}{{Igoshev} \&
  {Popov}}{2014}]{ip14}
{Igoshev} A.~P.,  {Popov} S.~B.,  2014, \mn@doi [MNRAS]
  {10.1093/mnras/stu1496}, \href
  {http://adsabs.harvard.edu/abs/2014MNRAS.444.1066I} {444, 1066}

\bibitem[\protect\citeauthoryear{{Igoshev} \& {Popov}}{{Igoshev} \&
  {Popov}}{2015}]{ip15}
{Igoshev} A.~P.,  {Popov} S.~B.,  2015, \mn@doi [Astronomische Nachrichten]
  {10.1002/asna.201512232}, \href
  {http://adsabs.harvard.edu/abs/2015AN....336..831I} {336, 831}

\bibitem[\protect\citeauthoryear{{Igoshev}, {Popov}  \& {Hollerbach}}{{Igoshev}
  et~al.}{2021}]{ip21}
{Igoshev} A.~P.,  {Popov} S.~B.,   {Hollerbach} R.,  2021, \mn@doi [Universe]
  {10.3390/universe7090351}, \href
  {https://ui.adsabs.harvard.edu/abs/2021Univ....7..351I} {7, 351}

\bibitem[\protect\citeauthoryear{{Igoshev}, {Frantsuzova}, {Gourgouliatos},
  {Tsichli}, {Konstantinou}  \& {Popov}}{{Igoshev} et~al.}{2022}]{Igoshev22}
{Igoshev} A.~P.,  {Frantsuzova} A.,  {Gourgouliatos} K.~N.,  {Tsichli} S.,
  {Konstantinou} L.,   {Popov} S.~B.,  2022, \mn@doi [\mnras]
  {10.1093/mnras/stac1648}, \href
  {https://ui.adsabs.harvard.edu/abs/2022MNRAS.514.4606I} {514, 4606}

\bibitem[\protect\citeauthoryear{{Johnston} \& {Kramer}}{{Johnston} \&
  {Kramer}}{2019}]{jk19}
{Johnston} S.,  {Kramer} M.,  2019, \mn@doi [\mnras] {10.1093/mnras/stz2865},
  \href {https://ui.adsabs.harvard.edu/abs/2019MNRAS.490.4565J} {490, 4565}

\bibitem[\protect\citeauthoryear{{Keith}, {Johnston}, {Weltevrede}  \&
  {Kramer}}{{Keith} et~al.}{2010}]{keith10}
{Keith} M.~J.,  {Johnston} S.,  {Weltevrede} P.,   {Kramer} M.,  2010, \mn@doi
  [\mnras] {10.1111/j.1365-2966.2009.15926.x}, \href
  {https://ui.adsabs.harvard.edu/abs/2010MNRAS.402..745K} {402, 745}

\bibitem[\protect\citeauthoryear{{Lyne} \& {Manchester}}{{Lyne} \&
  {Manchester}}{1988}]{lyne88}
{Lyne} A.~G.,  {Manchester} R.~N.,  1988, \mn@doi [MNRAS]
  {10.1093/mnras/234.3.477}, \href
  {http://adsabs.harvard.edu/abs/1988MNRAS.234..477L} {234, 477}

\bibitem[\protect\citeauthoryear{{Lyne}, {Manchester}  \& {Taylor}}{{Lyne}
  et~al.}{1985}]{lmt85}
{Lyne} A.~G.,  {Manchester} R.~N.,   {Taylor} J.~H.,  1985, \mn@doi [\mnras]
  {10.1093/mnras/213.3.613}, \href
  {https://ui.adsabs.harvard.edu/#abs/1985MNRAS.213..613L} {213, 613}

\bibitem[\protect\citeauthoryear{{Maciesiak}, {Gil}  \& {Ribeiro}}{{Maciesiak}
  et~al.}{2011}]{maciesiak11a}
{Maciesiak} K.,  {Gil} J.,   {Ribeiro} V. A.~R.~M.,  2011, \mn@doi [\mnras]
  {10.1111/j.1365-2966.2011.18471.x}, \href
  {https://ui.adsabs.harvard.edu/abs/2011MNRAS.414.1314M} {414, 1314}

\bibitem[\protect\citeauthoryear{{Malov} \& {Nikitina}}{{Malov} \&
  {Nikitina}}{2011}]{nikitina11}
{Malov} I.~F.,  {Nikitina} E.~B.,  2011, \mn@doi [Astronomy Reports]
  {10.1134/S1063772911100076}, \href
  {http://adsabs.harvard.edu/abs/2011ARep...55..878M} {55, 878}

\bibitem[\protect\citeauthoryear{{Malov} \& {Nikitina}}{{Malov} \&
  {Nikitina}}{2013}]{malov13}
{Malov} I.~F.,  {Nikitina} E.~B.,  2013, \mn@doi [Astronomy Reports]
  {10.1134/S106377291311005X}, \href
  {https://ui.adsabs.harvard.edu/abs/2013ARep...57..833M} {57, 833}

\bibitem[\protect\citeauthoryear{{Manchester}, {Hobbs}, {Teoh}  \&
  {Hobbs}}{{Manchester} et~al.}{2005}]{atnf}
{Manchester} R.~N.,  {Hobbs} G.~B.,  {Teoh} A.,   {Hobbs} M.,  2005, \mn@doi
  [AJ] {10.1086/428488}, \href
  {http://adsabs.harvard.edu/abs/2005AJ....129.1993M} {129, 1993}

\bibitem[\protect\citeauthoryear{{Manchester} et~al.,}{{Manchester}
  et~al.}{2010}]{manch10}
{Manchester} R.~N.,  et~al., 2010, \mn@doi [\apj]
  {10.1088/0004-637X/710/2/1694}, \href
  {https://ui.adsabs.harvard.edu/abs/2010ApJ...710.1694M} {710, 1694}

\bibitem[\protect\citeauthoryear{{Novoselov}, {Beskin}, {Galishnikova},
  {Rashkovetskyi}  \& {Biryukov}}{{Novoselov} et~al.}{2020}]{Novoselov2020}
{Novoselov} E.~M.,  {Beskin} V.~S.,  {Galishnikova} A.~K.,  {Rashkovetskyi}
  M.~M.,   {Biryukov} A.~V.,  2020, \mn@doi [\mnras] {10.1093/mnras/staa904},
  \href {https://ui.adsabs.harvard.edu/abs/2020MNRAS.494.3899N} {494, 3899}

\bibitem[\protect\citeauthoryear{{{\"O}zel} \& {Freire}}{{{\"O}zel} \&
  {Freire}}{2016}]{ozel16}
{{\"O}zel} F.,  {Freire} P.,  2016, \mn@doi [Ann. Rev. of Astronomy and
  Astrophysics] {10.1146/annurev-astro-081915-023322}, \href
  {http://adsabs.harvard.edu/abs/2016ARA%26A..54..401O} {54, 401}

\bibitem[\protect\citeauthoryear{{Philippov}, {Tchekhovskoy}  \&
  {Li}}{{Philippov} et~al.}{2014}]{phil14}
{Philippov} A.,  {Tchekhovskoy} A.,   {Li} J.~G.,  2014, \mn@doi [MNRAS]
  {10.1093/mnras/stu591}, \href
  {http://adsabs.harvard.edu/abs/2014MNRAS.441.1879P} {441, 1879}

\bibitem[\protect\citeauthoryear{{Popov}, {Pons}, {Miralles}, {Boldin}  \&
  {Posselt}}{{Popov} et~al.}{2010}]{popov10}
{Popov} S.~B.,  {Pons} J.~A.,  {Miralles} J.~A.,  {Boldin} P.~A.,   {Posselt}
  B.,  2010, \mn@doi [MNRAS] {10.1111/j.1365-2966.2009.15850.x}, \href
  {http://adsabs.harvard.edu/abs/2010MNRAS.401.2675P} {401, 2675}

\bibitem[\protect\citeauthoryear{{Radhakrishnan} \& {Cooke}}{{Radhakrishnan} \&
  {Cooke}}{1969}]{RVM}
{Radhakrishnan} V.,  {Cooke} D.~J.,  1969, \aplett, \href
  {https://ui.adsabs.harvard.edu/abs/1969ApL.....3..225R} {3, 225}

\bibitem[\protect\citeauthoryear{{Rankin}}{{Rankin}}{1990}]{rankin90}
{Rankin} J.~M.,  1990, \mn@doi [\apj] {10.1086/168530}, \href
  {https://ui.adsabs.harvard.edu/abs/1990ApJ...352..247R} {352, 247}

\bibitem[\protect\citeauthoryear{{Rankin}}{{Rankin}}{1993a}]{rankin93a}
{Rankin} J.~M.,  1993a, \mn@doi [\apjs] {10.1086/191758}, \href
  {http://adsabs.harvard.edu/abs/1993ApJS...85..145R} {85, 145}

\bibitem[\protect\citeauthoryear{{Rankin}}{{Rankin}}{1993b}]{rankin93b}
{Rankin} J.~M.,  1993b, \mn@doi [Astrophys. J.] {10.1086/172361}, \href
  {http://adsabs.harvard.edu/abs/1993ApJ...405..285R} {405, 285}

\bibitem[\protect\citeauthoryear{{Ridley} \& {Lorimer}}{{Ridley} \&
  {Lorimer}}{2010}]{ridley10}
{Ridley} J.~P.,  {Lorimer} D.~R.,  2010, \mn@doi [MNRAS]
  {10.1111/j.1365-2966.2010.16342.x}, \href
  {http://adsabs.harvard.edu/abs/2010MNRAS.404.1081R} {404, 1081}

\bibitem[\protect\citeauthoryear{{Spitkovsky}}{{Spitkovsky}}{2006}]{spitkovsky06}
{Spitkovsky} A.,  2006, \mn@doi [Astrophys. J. Lett.] {10.1086/507518}, \href
  {http://adsabs.harvard.edu/abs/2006ApJ...648L..51S} {648, L51}

\bibitem[\protect\citeauthoryear{{Tauris} \& {Manchester}}{{Tauris} \&
  {Manchester}}{1998}]{tm98}
{Tauris} T.~M.,  {Manchester} R.~N.,  1998, \mn@doi [MNRAS]
  {10.1046/j.1365-8711.1998.01369.x}, \href
  {http://adsabs.harvard.edu/abs/1998MNRAS.298..625T} {298, 625}

\bibitem[\protect\citeauthoryear{{Vigan{\`o}}, {Rea}, {Pons}, {Perna},
  {Aguilera}  \& {Miralles}}{{Vigan{\`o}} et~al.}{2013}]{vigano13}
{Vigan{\`o}} D.,  {Rea} N.,  {Pons} J.~A.,  {Perna} R.,  {Aguilera} D.~N.,
  {Miralles} J.~A.,  2013, \mn@doi [MNRAS] {10.1093/mnras/stt1008}, \href
  {http://adsabs.harvard.edu/abs/2013MNRAS.434..123V} {434, 123}

\bibitem[\protect\citeauthoryear{{Weltevrede} \& {Johnston}}{{Weltevrede} \&
  {Johnston}}{2008}]{welt08}
{Weltevrede} P.,  {Johnston} S.,  2008, \mn@doi [MNRAS]
  {10.1111/j.1365-2966.2008.13382.x}, \href
  {http://adsabs.harvard.edu/abs/2008MNRAS.387.1755W} {387, 1755}

\makeatother
\end{thebibliography}

.






\bsp	
\label{lastpage}
\end{document}